\documentclass[letterpaper,twocolumn,10pt]{article}
\usepackage{usenix2019}

\usepackage{booktabs}  
\usepackage{array}
\usepackage{multirow}

\usepackage{graphicx}
\usepackage{subcaption}
\usepackage{epsfig}

\usepackage{mathtools}
\usepackage{amsmath}

\usepackage{amssymb}
\usepackage{algorithm}
\usepackage{algorithmicx}
\usepackage[noend]{algpseudocode}

\usepackage[hyphens]{url}

\usepackage{color}
\usepackage{xspace}
\usepackage{ifthen}
\usepackage{balance}

\newcommand{\ie}{{\em i.e.}}
\newcommand{\eg}{{\em e.g.}}

\newcommand{\name}{Elmo\xspace}

\newcommand{\showComments}{yes}

\newcommand{\note}[2]{
    \ifthenelse{\equal{\showComments}{yes}}{\textcolor{#1}{#2}}{}
}

\newcommand{\ms}[1]{\note{blue}{Shahbaz: #1}}

\newcommand{\supsym}[1]{\raisebox{4pt}{{\footnotesize #1}}}
\newcommand{\pu}{\supsym{$\star$}}
\newcommand{\vm}{\supsym{$\dag$}}
\newcommand{\ti}{\supsym{$\ddag$}}

\usepackage[nocompress]{cite}

\begin{document}

\title{\name{}: Source-Routed Multicast for Cloud Services}


\author{
{\rm Muhammad Shahbaz\pu, Lalith Suresh\vm, Jen Rexford\pu, Nick Feamster\pu, 
Ori Rottenstreich\ti, and Mukesh Hira\vm}\\
\pu\normalsize{Princeton
University}~~~\vm\normalsize{VMware}~~~\ti\normalsize{Technion}
} 


\maketitle


\begin{abstract}
We present \name{}, a system that addresses the multicast scalability problem
in multi-tenant data centers. Modern cloud applications frequently
exhibit one-to-many communication patterns and, at the same time, require
sub-millisecond latencies and high throughput. IP multicast can achieve these
requirements but has control- and data-plane scalability limitations that make
it challenging to offer it as {\em a service for hundreds of thousands of
tenants}, typical of cloud environments. Tenants, therefore, must rely on
unicast-based approaches (\eg, application-layer or overlay-based) to support
multicast in their applications, imposing overhead on throughput and end host
CPU utilization, with higher and unpredictable latencies. 

\name{} scales network multicast by taking advantage of emerging programmable
switches and the unique characteristics of data-center networks; specifically,
the symmetric topology and short paths in a data center. Elmo encodes
multicast group information inside packets themselves, reducing the need to
store the same information in network switches. In a three-tier data-center
topology with 27K hosts, \name{} supports a million multicast groups using a
325-byte packet header, requiring as few as 1.1K multicast group-table entries
on average in leaf switches, with a traffic overhead as low as 5\% over ideal
multicast.
\end{abstract}

\section{Introduction}
\label{sec:introduction}

Cloud applications are commonly driven by systems that deliver large amounts of
data to \emph{groups} of endpoints in a multi-tenant data center. Common
workloads include streaming telemetry
workloads~\cite{azure-telemetry-reporting, massie2004ganglia, rfc3176,
openconfig-streaming-telemetry}, where hosts continuously send telemetry data
in incremental updates to a set of collectors; replication for
databases~\cite{galeracluster} and
state-machines~\cite{Ports:2015:DDS:2789770.2789774,
Lamport:1998:PP:279227.279229, lamport2006fast}; distributed
messaging~\cite{jgroups-multicast, akka-udp},
file-sharing~\cite{apache-hadoop}, and machine
learning~\cite{Mai:2015:ONP:2827719.2827721} frameworks; as well as schedulers
and load balancers that require a steady stream of telemetry information about
the load on servers to make effective server selection
decisions~\cite{Suresh:2015:CCT:2789770.2789806, elastic-discovery-plugins}.
Publish-subscribe systems are also common building blocks for large-scale
systems today~\cite{google-pubsub, Garg:2015:LAK:2800214, hintjens2013zeromq,
videla2012rabbitmq}. These systems create a large number of publish-subscribe
topics per tenant~\cite{google-pubsub-quotas}. Infrastructure
applications~\cite{vmware-nsx} running on top of a provider's network,
likewise, need to replicate broadcast, unknown unicast, and multicast traffic
for its tenants~\cite{vxlan-on-nsx}. 

These types of workloads naturally suggest the use of multicast, yet today's
multi-tenant data centers typically do not deploy IP multicast~\cite{aws-faq,
gcp-faq, azure-faq}. In practice, IP multicast is not effective, since
data-center tenants introduce churn in the multicast state (\eg, due to virtual
machine allocation~\cite{Armbrust:2010:VCC:1721654.1721672, 202608} and
migration~\cite{Cully:2008:RHA:1387589.1387601,
Clark:2005:LMV:1251203.1251223}). IGMP~\cite{rfc3376} and PIM~\cite{rfc4601,
rfc5015, rfc4607} trigger many control messages during churn, querying the
entire PIM broadcast domain periodically; and are not robust to network
failures~\cite{Diot:2000:DII:2329003.2329353, rfc7431,
Wang:2000:IMF:850937.852509, Li:2013:SIM:2535372.2535380}. While SDN-based
solutions (like OpenFlow~\cite{openflow}) alleviate the control plane
shortcomings of IGMP and PIM, they still do not scale to support a large number
of groups in a multi-tenant data center. In particular, switching hardware
supports a limited number of group-table entries only, typically thousands to a
few tens of thousands~\cite{hardware-group-limit, multicast-group-capacity,
multicast-sdn-vxlan, Li:2013:SIM:2535372.2535380}. Cloud providers, however,
host hundreds of thousands of tenants~\cite{aws-users}, each of which may run
tens to hundreds of applications that might benefit from network multicast. If
cloud providers want to offer network multicast as a service to tenants, we
believe they need to scale up to \emph{millions} of multicast groups in a
single data center. 

In the absence of IP multicast, cloud providers and tenants typically
implement multicast using a unicast
overlay~\cite{Jannotti:2000:ORM:1251229.1251243, Castro:2003:SHM:945445.945474,
Banerjee:2002:SAL:633025.633045, Das:2002:SSW:647883.738420,
Vigfusson:2010:DMR:1755913.1755949, amazon-overlay-multicast,
weave-overlay-multicast}, which imposes load on the end host CPU and,
therefore, cannot match network line rates. Due to such scaling limitations,
certain classes of workloads (\eg, many workloads introduced by financial
applications~\cite{arista-financial-applications}) and approaches to scaling
existing workloads (like Coded MapReduce~\cite{li2015coded}) cannot benefit
from today's cloud-based infrastructure at all. 

In this work, we present \name{}, a system for network multicast that scales to
support millions of groups in a multi-tenant data center. Our approach to
scaling multicast groups is to encode the forwarding policy (\ie, multicast
tree) in the packet header, as opposed to maintaining group-table entries
inside network switches. Such a solution is more flexible and dynamic: the
group membership is encoded in the packet itself and groups can be reconfigured
by merely changing the information in the header of each packet. The challenge
in doing so involves finding the right balance between how much of the
forwarding information to place in the packet header (inflating both the
packet size and the complexity of parsing the packet header at each switch) and
how much state to put in each switch (increasing memory requirements at the
switch and limiting the rates at which multicast group memberships change). As
long as group sizes remain small enough to encode the entire multicast tree in
the packet, there is practically no limit on the number of groups \name{} can
support. To enable our scheme, we introduce a hardware primitive that is
inexpensive to implement in today's programmable switching ASICs.

Previous network multicast designs explored the tradeoff between packet header
size and switch memory in the context of arbitrarily flexible switch
architectures~\cite{Jokela:2009:LLS:1592568.1592592,
Ratnasamy:2006:RIM:1159913.1159917, rfc8279}. This paper, on the other hand,
studies this tradeoff in the context of multi-tenant data centers, where we can
exploit the characteristics of data-center topologies to design a more
efficient packet-header encoding that can be implemented and deployed on
programmable switches {\em today}. This new context allows us to
take advantage of the unique characteristics of data-center topologies. First,
data-center topologies tend to be symmetric (\ie, having core, spine, and leaf
tiers). Second, they have a limited number of switches on any individual path. 
The main result of this work is a system, \name{}, to encode multicast
forwarding policies in packets which takes advantage of these unique
characteristics of data-center topologies and to create an encoding for
multicast groups that is compact enough to fit in a header that can be parsed
by programmable switches that are being deployed in today's data
centers~\cite{Bosshart:2013:FMF:2486001.2486011, barefoot-tofino, xpliant}.

This paper makes the following contributions. First, we develop a technique for
compactly encoding multicast groups that are subtrees of multi-rooted Clos
topologies~(\S\ref{sec:encoding}), the prevailing topology for today's data
centers~\cite{facebook-fabric, NiranjanMysore:2009:PSF:1592568.1592575,
Greenberg:2009:VSF:1592568.1592576}. These topologies create an opportunity to
design a multicast group encoding that is compact enough for today's
programmable switches to process. Second, we optimize the encoding so that it
can be efficiently implemented in both hardware and software
targets~(\S\ref{sec:switches}); our hardware primitive requires only 0.0515\%
in additional area cost on a modern programmable switching ASIC. Our evaluation
shows that our encoding facilitates a feasible implementation in today's
multi-tenant data centers~(\S\ref{sec:evaluation}). In a data center with 27K
hosts, \name{} scales to millions of multicast groups with minimal group-table
entries and control-plane update overhead on switches. Lastly, \name{} supports
applications that use multicast without modification; we demonstrate two
such applications.

\section{\name{} Architecture}
\label{sec:architecture}

\begin{figure}[t!]
\centering  
\includegraphics[width=0.80\linewidth]{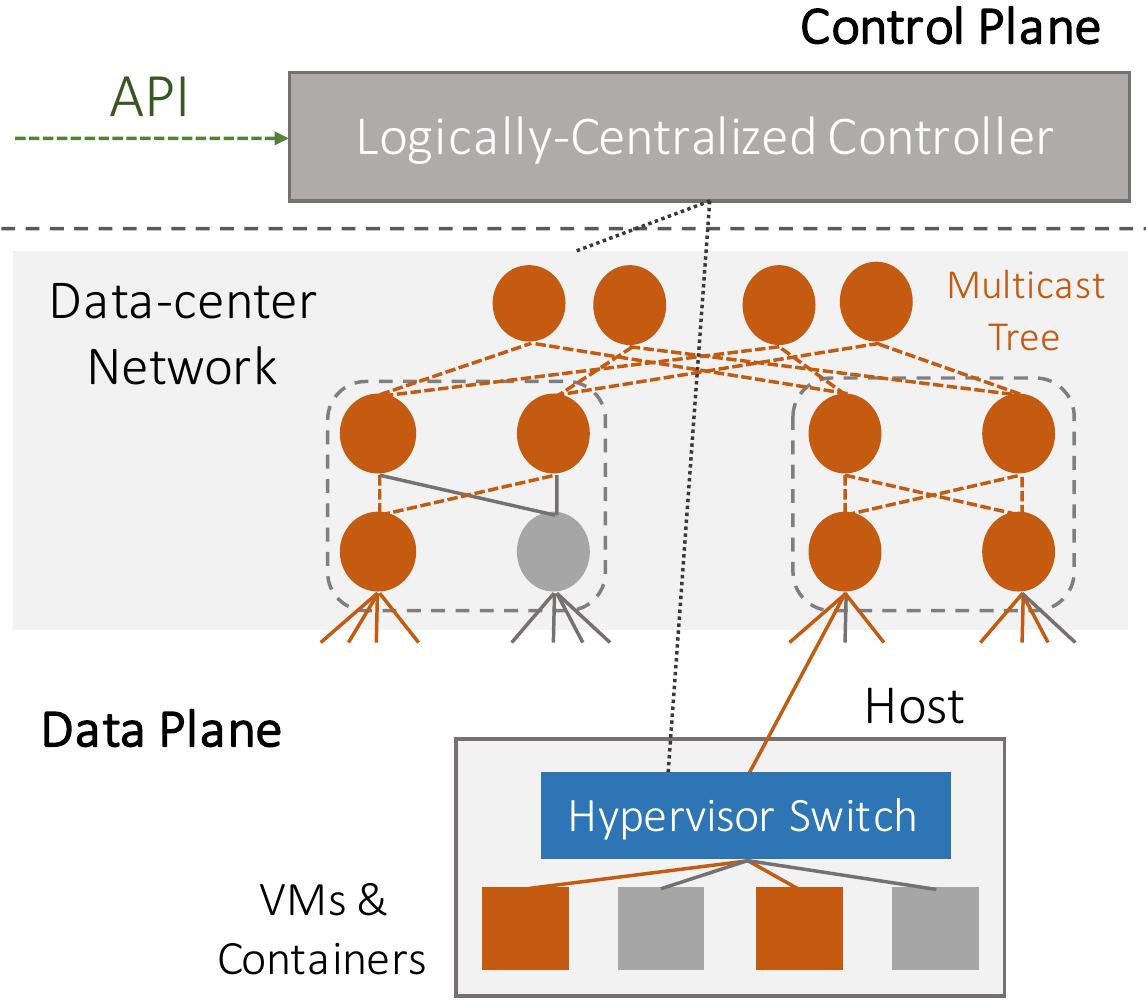}
\caption{\name{} architecture (with an example multicast tree, in orange).}
\label{fig:architecture}
\end{figure}

In \name{}, a {\em logically-centralized controller} manages multicast groups
for tenants by installing flow rules in \emph{hypervisor switches} (to
encapsulate packets with a compact encoding of the forwarding policy) and the
\emph{network switches} (to handle forwarding decisions for groups too large to
encode entirely in the packet header). Performing control-plane operations at
the controller and having the hypervisor switches place forwarding rules in
packet headers, significantly reduces the burden on network switches for
handling a large number of multicast groups. Figure~\ref{fig:architecture}
summarizes our architecture.

\paragraph{Logically-centralized controller.} The logically-centralized
controller receives join and leave requests for multicast groups via an
application programming interface (API). Cloud providers already expose such
APIs~\cite{cloud-apis} for tenants to request VMs, load balancers, firewalls,
and other services. Each multicast group consists of a set of tenant VMs. The
controller knows the physical location of each tenant VM, as well as the
current network topology---including the capabilities and capacities of the
switches, along with unique identifiers for addressing these switches. Today's
data centers already maintain such soft state about network configuration at
the controller~\cite{NiranjanMysore:2009:PSF:1592568.1592575} (using
distributed directory systems~\cite{Greenberg:2009:VSF:1592568.1592576}). The
controller relies on a high-level language (like P4~\cite{p4,
Budiu:2017:PPL:3139645.3139648}) to configure the programmable switches at boot
time so that the switches can parse and process \name{}'s multicast packets.
The controller computes the multicast trees for each group and uses a control
interface (like P4Runtime~\cite{p4runtime}) to install match-action rules in
the switches at run time. When notified of events (\eg, network failures, and
group membership changes), the controller computes new rules and updates only
the affected switches (data-center controllers are capable of executing these
steps in sub-second timescales~\cite{NiranjanMysore:2009:PSF:1592568.1592575}.)
The controller uses a clustering algorithm for computing compact encodings of
the multicast forwarding policies in packet headers (\S\ref{ssec:algorithm}).

\paragraph{Hypervisor switch.} A software
switch~\cite{Pfaff:2015:DIO:2789770.2789779, firestone2017vfp,
Shahbaz:2016:PPP:2934872.2934886}, running inside the hypervisor, intercepts
multicast data packets originating from VMs. The hypervisor switch matches the
destination IP address of a multicast group in the flow table to determine what
actions to perform on the packet. The actions determine: (i) where to forward
the packet and (ii) what header to push on the packet. The header consists of a
list of rules (packet rules, or {\em p}-rules for short)---each containing a
set of output ports along with zero or more switch identifiers---that
intermediate network switches use to forward the packet. These {\em p}-rules
encode the multicast tree of a given group inside the packet, obviating the
need for network switches to store a large number of multicast forwarding rules
or update these rules when the tree changes. Hypervisor switches run as
software at the edge of the network, they do not have the hard constraints on
flow-table sizes and rule update frequency that network switches
have~\cite{ovs-perf, He:2015:MCP:2774993.2775069,
Pfaff:2015:DIO:2789770.2789779, firestone2017vfp}. Each hypervisor switch only
maintains flow rules for multicast groups that have member VMs running on the
same host, discarding packets belonging to other groups.

\paragraph{Network switch.} Upon receiving a multicast data packet, a physical
switch (or network switch) running inside the network simply parses the header
to look for a matching {\em p}-rule (\ie, a {\em p}-rule containing the
switch's own identifier) and forwards the packet to the associated output
ports, as well as popping {\em p}-rules when they are no longer needed to save
bandwidth. When a multicast tree is too large to encode entirely in the packet
header, a network switch may have its own group-table rule (called a switch
rule, or {\em s}-rule for short). As such, if a packet header contains no
matching {\em p}-rule, the network switch checks for an {\em s}-rule matching
the destination IP address (multicast group) and forwards the packet
accordingly. If no matching {\em s}-rule exists, the network switch forwards
the packet based on a default {\em p}-rule---the last {\em p}-rule in the
packet header. Elmo installs only a small number of {\em s}-rules on network
switches, consistent with the small group tables available in high-speed
hardware switches~\cite{hardware-group-limit, multicast-group-capacity,
multicast-sdn-vxlan}. The network switches in data centers form a tiered
topology (\eg, Clos) with leaf and spine switches grouped into pods, and core
switches. Together they enable \name{} to encode multicast trees efficiently.

\section{Encoding Multicast Trees}
\label{sec:encoding}

Upon receiving a multicast data packet, a switch must identify what set of
output ports (if any) to forward the packet while ensuring it is sent on every
output port in the tree and as few extra ports as possible. In this section, we
first describe how to represent multicast trees efficiently, by capitalizing on
the structure of data centers (topology and short paths) and capabilities of
programmable switches (flexible parsing and forwarding).

\begin{figure}[t!]
\centering  
\includegraphics[width=0.85\linewidth]{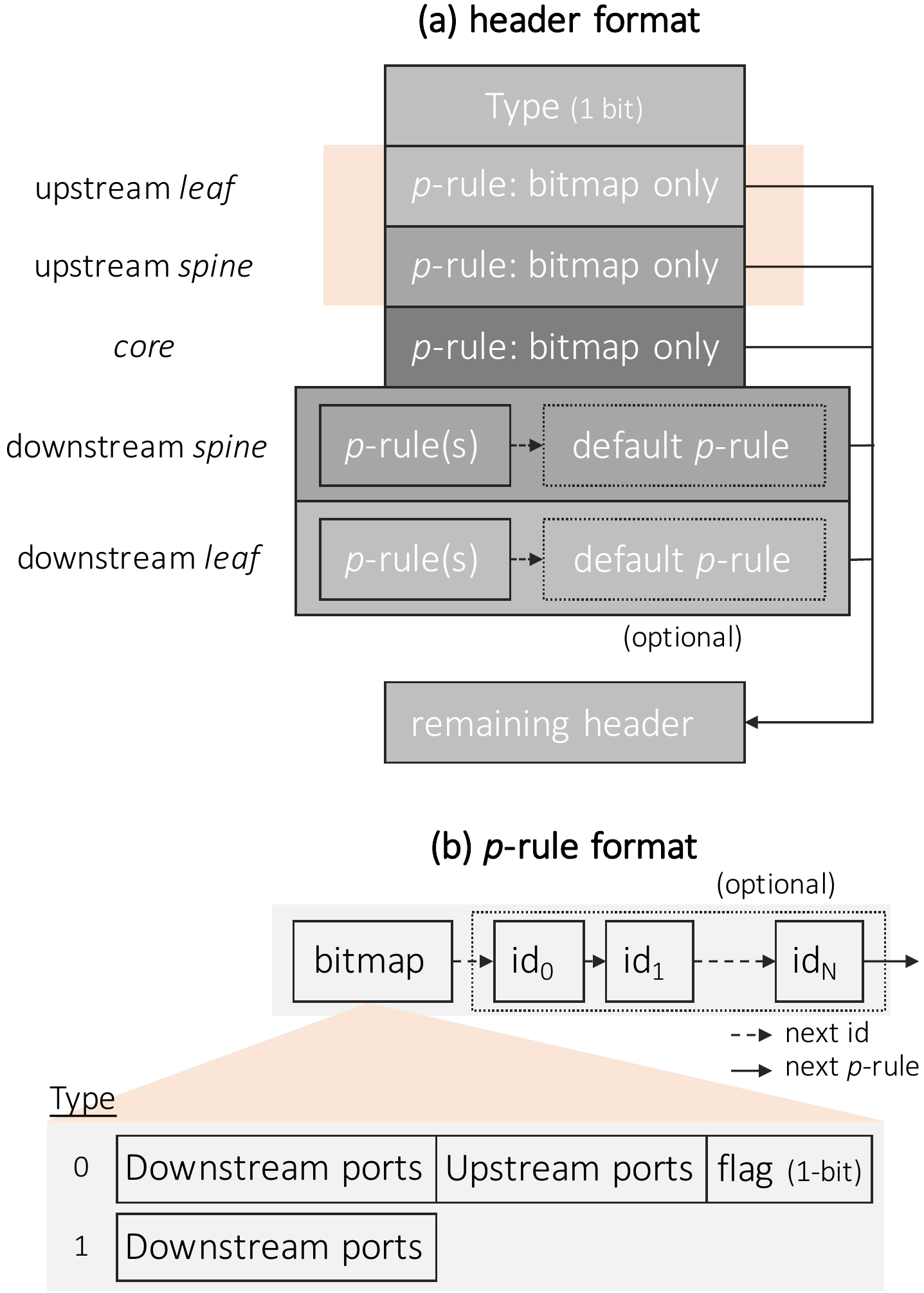}
\caption{\name{}'s header and {\em p}-rule format.}
\label{fig:header-format}
\end{figure}

\subsection{Packet Header Design}
\label{ssec:compact-packet-headers}

\name{} encodes a multicast forwarding policy efficiently in a packet header
as a list of {\em p}-rules (Figure~\ref{fig:header-format}a). 
We introduce five key design decisions (\textbf{{\em D1-5}}) that make our
{\em p}-rule encoding both {\em compact} and {\em simple} for switches to
process.

\begin{figure*}[t!]
\centering  
\includegraphics[width=\linewidth]
{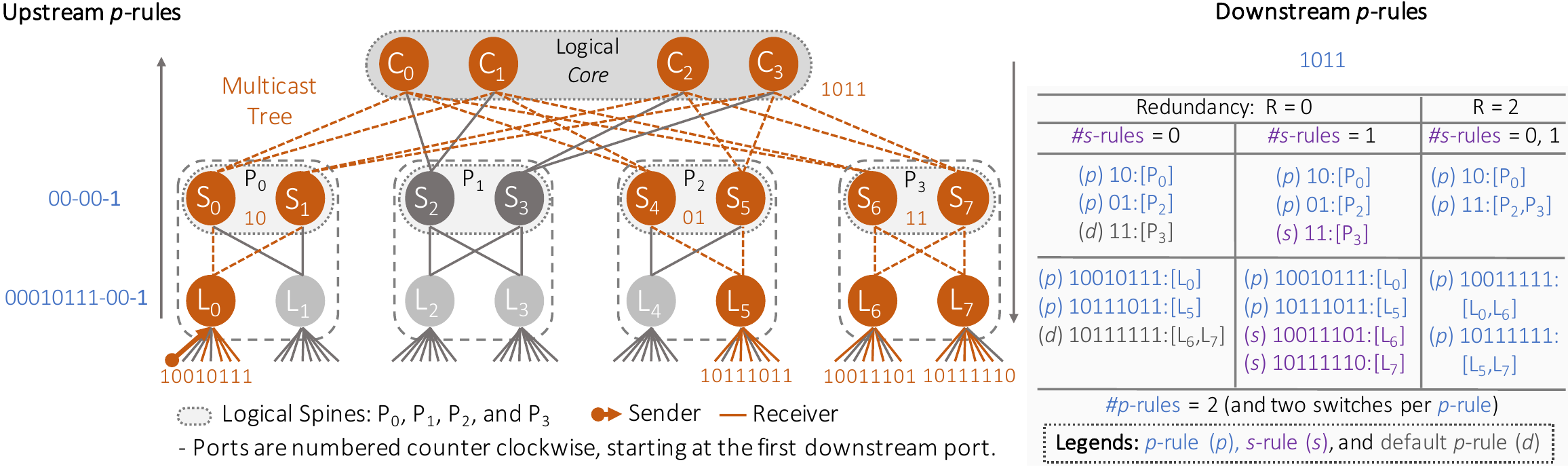}
\caption{An example multicast tree on a three-tier multi-rooted Clos topology
with upstream and downstream {\em p}- and {\em s}-rules assignment for a
group. A packet originating from the sender is forwarded up to the logical core
using the upstream {\em p}-rules, and down to the receivers using the
downstream {\em p}-rules (and {\em s}-rules). For example, when $R=0$ and
\#{\em s}-rules $=$ 1, a packet arriving at $P_2$ ($S_4$ or $S_5$) from the
core is forwarded using the {\em p}-rule 01, whereas at $P_3$, it is forwarded
using the {\em s}-rule 11.}
\label{fig:clos-topology-multicast-tree}
\end{figure*}

Throughout this section, we use a three-tier multi-rooted Clos topology
(Figure~\ref{fig:clos-topology-multicast-tree}) with a multicast group
stretching across three pods (marked in orange) as a running example. The
topology consists of four core switches and pods, and two spine and leaf
switches per pod. Each leaf switch further connects to eight hosts.

\paragraph{D1: Encoding switch output ports in a bitmap.}
Each {\em p}-rule uses a simple bitmap to represent the set of switch output
ports (typically, 48 ports) that should forward the packet
(Figure~\ref{fig:header-format}b). Using a bitmap is desirable because it is
the internal data structure that network switches use
to direct a packet to multiple output
ports~\cite{Bosshart:2013:FMF:2486001.2486011}. 
Alternative encoding strategies use destination group members, encoded as bit
strings~\cite{rfc8279}; bloom filters, representing link
memberships~\cite{Jokela:2009:LLS:1592568.1592592,
Ratnasamy:2006:RIM:1159913.1159917}; or simply a lists of IP
addresses~\cite{boivie2000small} to identify the set of output ports. However,
these representations cannot be efficiently processed by network switches
without violating line-rate guarantees
(discussed in detail in \S\ref{sec:related-work}). 

Having a separate {\em p}-rule---with a bitmap and an identifier for each
switch---for the multicast group in our example three-tier Clos topology
(Figure~\ref{fig:clos-topology-multicast-tree}) needs a header size of
161~bits. For identifiers, we use two bits to identify the four core switches
and three bits for spine and leaf switches, each.

With bitmap encoding, the {\em p}-rule for switch $L_0$ (in
Figure~\ref{fig:clos-topology-multicast-tree}) may look like
\verb|00010111-00|:$L_0$. Each bit corresponds to an output port on the given
switch, indicating the downstream and upstream ports participating in the
multicast tree.

\paragraph{D2: Encoding on the logical topology.}
Instead of having separate {\em p}-rules for each switch in the multicast
tree, \name{} exploits the tiered architecture and short path
lengths\footnote{\eg, a maximum of five hops in the Facebook Fabric
topology~\cite{facebook-fabric}} in today's data-center topologies to reduce
the number of required {\em p}-rules. In multi-rooted Clos topologies, such as
our example topology (Figure~\ref{fig:clos-topology-multicast-tree}),
leaf-to-spine and spine-to-core links use multipathing. All spine switches in
the same pod behave as one giant logical switch (forwarding packets to the
same destination leaf switches), and all core switches together behave as one
logical core switch (forwarding packets to the same pods). We refer to the
\emph{logical topology} as one where there is a single logical spine switch per
pod, and a single logical core switch connected to pods.

We order {\em p}-rules inside a packet by layers according to the following
topological ordering: {\em upstream leaf, upstream spine, core, downstream
spine, and downstream leaf} (Figure~\ref{fig:header-format}a). Doing so
also accounts for varying switch port densities per layer. Organizing {\em
p}-rules by layers together with the other characteristics of the logical
topology allow us to further reduce header size and traffic overhead of a
multicast group in four key ways:

\paragraph{a)} We only require one {\em p}-rule per \emph{logical}
switch, with all switches belonging to the same logical group using not only
the same bitmap to send packets to output ports, but also requiring only one
logical switch identifier in the {\em p}-rule.

\paragraph{b)} A multicast packet visits a layer only once, both in its
upstream and downstream path. Grouping {\em p}-rules by layer, therefore,
allows switches to pop all headers of that layer when forwarding a packet from
one layer to another. This is because {\em p}-rules from any given layer are
irrelevant to subsequent layers in the path. This also exploits the capability
of programmable switches to decapsulate headers at line rate, discussed
in~\S\ref{sec:switches}. Doing so further reduces traffic overhead. 

\paragraph{c)} For switches in the upstream path, {\em p}-rules contain
only the bitmap---including the downstream and upstream ports, and a {\em flag}
bit (Figure~\ref{fig:header-format}b: $Type=0$)---without a switch identifier
list. These switches simply read the first upstream {\em p}-rule in the packet
header (Figure~\ref{fig:header-format}a), popping it before forwarding to the
next switch in the upstream path. The flag bit indicates whether a switch
should use the configured underlying multipath scheme (\eg, ECMP,
CONGA~\cite{Alizadeh:2014:CDC:2619239.2626316}, or
HULA~\cite{Katta:2016:HSL:2890955.2890968}) or not. Otherwise, the
upstream ports are used for forwarding packets upward to multiple switches
in cases where no single spine or core has connectivity to all members of a
multicast group (\eg, due to network failures,~\S\ref{sub:upstream-ports}).

\paragraph{d)} Again, because a multicast packet visits a layer only once, the
only switches that require upstream ports represented in their bitmaps are the
leaf and spine switches in the upstream path (Figure~\ref{fig:header-format}b:
$Type=0$). The bitmaps of all other switches only require their downstream
ports represented using bitmaps (Figure~\ref{fig:header-format}b: $Type=1$).
The shorter bitmaps for these switches, therefore, reduce space usage even
further. Note, upstream ports differ based on the source, whereas, downstream
ports remain identical within the same multicast group.

In our example (Figure~\ref{fig:clos-topology-multicast-tree}), encoding on the
logical topology drops the header size to 83 bits (a reduction of 48.44\% from
{\em D1}).

\paragraph{D3: Sharing a bitmap across switches.} Even with a logical topology,
having a separate {\em p}-rule for each switch in the downstream path could
lead to very large packet headers, imposing bandwidth overhead on the network.
In addition, network switches have restrictions on the packet header sizes they
can parse (\eg, 512 bytes~\cite{Bosshart:2013:FMF:2486001.2486011}), limiting
the number of {\em p}-rules we can encode in each packet. To further reduce
header sizes, \name{} assigns multiple switches within each layer (enabled by
{\em D2}) to the same {\em p}-rule, if the switches have the same---or
similar---bitmaps. Mapping multiple switches to a single bitmap, as a bitwise
OR of their individual bitmap, reduces header sizes because the output bitmap
of a rule requires more bits to represent than switch identifiers; for example,
a datacenter with 27K hosts has approximately 1000 switches (needing 10 bits to
represent switch identifiers), whereas switch port densities range from 48 to
576 (requiring that many bits)~\cite{Alizadeh:2014:CDC:2619239.2626316}. The
algorithm to identify sets of switches with similar bitmaps is described in
\S\ref{ssec:algorithm}. 

We encode the set of switches as a simple \emph{list} of switch identifiers, as
shown in Figure~\ref{fig:header-format}b. Alternate encodings, such as bloom
filters~\cite{Bloom:1970:STH:362686.362692}, are more complicated to
implement---requiring a switch to account for false positives, where multiple
{\em p}-rules are a ``match.'' To keep false-positive rates manageable, these
approaches lead to large filters~\cite{Li:2013:SIM:2535372.2535380}, which is
less efficient than having a list, as the number of switches with similar
bitmaps is relatively small compared to the total number of switches in the
data-center network.

With {\em p}-rule sharing, such that the bitmaps of assigned switches differ by
at most two bits (\ie, $R=2$, \S\ref{ssec:algorithm}), logical switches
$P_2$ and $P_3$ (in Figure~\ref{fig:clos-topology-multicast-tree}) share a 
downstream {\em p}-rule at the spine layer. At the leaf layer, $L_0$ shares a
downstream {\em p}-rule with $L_6$ and $L_5$ with $L_7$. This further brings
down the header size to 62 bits (a decrease of 25.30\% from {\em D2}).

\paragraph{D4: Limiting header size using default p-rules.}
A default {\em p}-rule accommodates all switches that do not share a {\em
p}-rule with other switches ({\em D3}). Default {\em p}-rules act as a
mechanism to limit the total number of {\em p}-rules in the header.  For
example, in Figure~\ref{fig:clos-topology-multicast-tree}, with $R=0$ and no
{\em s}-rules, leaf switches $L_6$ and $L_7$ both get assigned to a default
{\em p}-rule. The default {\em p}-rules are analogous to the lowest priority
rule in the context of a flow table. They are appended after all the other {\em
p}-rules of a downstream layer in the header (Figure~\ref{fig:header-format}a).

The output bitmap for a default {\em p}-rule is computed as the bitwise OR
of port memberships of all switches being mapped to the default rule.
In the limiting case, the default {\em p}-rule causes a packet to be forwarded
out of all output ports connected to the next layer at a switch (packets
\emph{only} make progress to the destination hosts); thereby, increasing
traffic overhead because of the extra transmissions.

\paragraph{D5: Reducing traffic overhead using s-rules.}
Combining all the techniques discussed so far allows \name{} to represent any
multicast tree without using \emph{any} state in the network switches. This is
made possible because of the default {\em p}-rules, which accommodate any
switches not captured by other {\em p}-rules. However, the use of the default
{\em p}-rule (and bitmap sharing across switches) results in extra packet
transmissions that increase traffic overhead.

To reduce the traffic overhead without increasing header size, we exploit the
fact that switches already support multicast group tables. Each entry, an
{\em s}-rule, in the group table matches a multicast group identifier and
sends a packet out on multiple ports. Before assigning a switch to a default
{\em p}-rule for a multicast group, we first check if the switch has space for
an {\em s}-rule. If so, we install an {\em s}-rule in that switch, and assign
only those switches to the default {\em p}-rule that have no spare {\em s}-rule
capacity. For example, in Figure~\ref{fig:clos-topology-multicast-tree}, with
{\em s}-rule capacity of one entry per switch (and $R=0$), both
leaf switches $L_6$ and $L_7$ now have an {\em s}-rule entry instead of the
default {\em p}-rule, as in the previous case ({\em D4}). 

\subsection{Algorithm for Generating {\em p}- and {\em s}-Rules}
\label{ssec:algorithm}
Having discussed the mechanisms of our design, we now explain how \name{}
expresses a group's multicast tree as a combination of {\em p}- and
{\em s}-rules. The algorithm is executed once per downstream layer for each
group. The input to the algorithm is a set of switch identifiers and their
output ports for a multicast tree (\emph{input bitmaps}).

\paragraph{Constraints.}
Every layer needs its own {\em p}-rules. Within each layer, we ensure that no
more than $H_{max}$ {\em p}-rules are used. We budget a separate $H_{max}$ per
layer such that the total number of {\em p}-rules is within a header size
limit. This is straightforward to compute because (i) we bound the number of
switches per {\em p}-rule to $K_{max}$---restricting arbitrary number of
switches from sharing a {\em p}-rule and inflating the header size---so the
maximum size of each {\em p}-rule is known a priori, and (ii) the number of
{\em p}-rules required in the upstream direction is known, leaving only the
downstream spine and leaf switches. Of these, downstream leaf switches use most
of the header capacity (\S\ref{sec:evaluation}).

A network switch has space for at most $F_{max}$ {\em s}-rules, a shared 
resource across all multicast groups. For {\em p}-rule sharing, we identify
groups of switches to share an output bitmap where the bitmap is the bitwise OR
of all the input bitmaps. To reduce traffic overhead, we bound the total number
of spurious transmissions resulting from a shared {\em p}-rule to $R$, where
$R$ is computed as the sum of Hamming Distances of each input bitmap to the
output bitmap.

\begin{algorithm}[t]
{
\small
\caption{Clustering algorithm for each layer of a group}
\label{alg:cluster}
\begin{algorithmic}[1]
\Statex \textbf{Constants:} $R$, $H_{max}$, $K_{max}$, $F_{max}$
\Statex \textbf{Inputs:} Set of all switches $S$, Bitmaps $B = b_{i} \forall i
\in S$
\Statex \textbf{Outputs:} {\em p}-rules, {\em s}-rules and default-{\em p}-rule
\State {\em p}-rules $\gets \varnothing$, {\em s}-rules $\gets \varnothing$,
default-{\em p}-rule $\gets \varnothing$
\State unassigned $\gets$ $B$, $K \gets K_{max}$
\While{unassigned $\ne \varnothing$ and $|${\em p}-rules$|$ $<H_{max}$}
\label{pRuleStart}
\State bitmaps $\gets$ approx-min-k-union($K$, unassigned)
\label{minK}
\State output-bm $\gets$ Bitwise OR of all $b_i \in$ bitmaps
\label{outputBitm}
\If{dist($b_i$, output-bm) $\le R$ $\forall ~b_i\in$ bitmaps}
\label{Kconstraint}
\State {\em p}-rules $\gets$ {\em p}-rules $\cup~$ bitmaps
\label{pRuleAssign}
\State unassigned $\gets$ unassigned $\setminus$ bitmaps
\label{pRuleRmv}
\Else
\State $K \gets K - 1$
\label{smallGrps}
\EndIf
\EndWhile
\label{pRuleEnd}
\ForAll{$b_i \in~$ unassigned} 
\If{switch $i$ has $|${\em s}-rules$|$ $<F_{max}$}
\State {\em s}-rules $\gets$ {\em s}-rules $ \cup ~\{b_i\}$
\label{sRule}
\Else
\State default-{\em p}-rule $\gets$ default-{\em p}-rule $ \cup ~\{b_i\}$
\label{defRule}
\EndIf
\EndFor
\Return {\em p}-rules, {\em s}-rules, default-{\em p}-rule
\end{algorithmic}
}
\end{algorithm}

\paragraph{Clustering algorithm.} The problem of determining which switches
share a {\em p}-rule maps to a well-known MIN-K-UNION problem, which is NP-hard
but has approximate variants available~\cite{vinterbo2002note}. Given the set of
bitmaps $B=\{b_1, b_2, \dots, b_n\}$, the goal is to find $K$ sets such that
the cardinality of their union is minimized. In our case, a set is a
bitmap---indicating the presence or absence of a switch port in a multicast
tree---and the goal is to find $K$ such bitmaps whose bitwise OR yields the
minimum number of set bits.

Algorithm~\ref{alg:cluster} shows our solution. For each group, we assign {\em
p}-rules until $H_{max}$ {\em p}-rules are assigned or all switches have been
assigned {\em p}-rules (Line~\ref{pRuleStart}). For {\em p}-rule sharing, we
apply an approximate MIN-K-UNION algorithm to find a group of $K$ input bitmaps
(Line~\ref{minK})~\cite{vinterbo2002note}. We then compute the bitwise OR of
these $K$ bitmaps to compute the resulting output bitmap
(Line~\ref{outputBitm}). If the output bitmap satisfies the traffic overhead
constraint (Line~\ref{Kconstraint}), we assign the $K$ switches to a {\em
p}-rule (Line~\ref{pRuleAssign}) and remove them from the set of unassigned
switches (Line~\ref{pRuleRmv}), and continue at Line~\ref{pRuleStart}.
Otherwise, we decrement $K$ and try to find smaller groups
(Line~\ref{smallGrps}). When $K=1$, any unassigned switches receive a {\em
p}-rule each. At any point if we encounter the $H_{max}$ constraint, we
fallback to computing {\em s}-rules for any remaining switches
(Line~\ref{sRule}). If the switches do not have any {\em s}-rule capacity left,
they are mapped to the default {\em p}-rule (Line~\ref{defRule}).

\subsection{Ensuring Reachability via Upstream Ports under Network Failures}
\label{sub:upstream-ports} 
Network failures (due to faulty switches or links) require recomputing
upstream {\em p}-rules for any affected groups. These rules are specific to
each source and, therefore, can either be computed by the controller or,
locally, at the hypervisor switches---which can scale and adapt more quickly to
failures using host-based techniques like
Clove~\cite{Katta:2017:CCL:3143361.3143401}. 

When a failure happens, a packet may not reach some members of a group via any
spine or core network switches using the underlying multipath scheme. In this
scenario, the controller deactivates multipathing using a flag bit~({\em D2})
and does not require updating the network switches. The controller disables the
flag bit in the bitmap of the upstream {\em p}-rules of the affected groups,
and forwards packets using the upstream ports. Furthermore, to identify the set
of possible paths that cover all members of a group, we reuse the same greedy
set-cover technique as used by
Portland~\cite{NiranjanMysore:2009:PSF:1592568.1592575} and therefore do not
expand on it in this paper; for a multicast group $G$, upstream ports in the
bitmap are set to forward packets to one or more spines (and cores) such that
the union of reachable hosts from the spine (and core) network switches covers
all the recipients of $G$. We evaluate how \name{} performs under failures in
\S\ref{sssec:update-overhead}.

\section{\name{} on Programmable Switches}
\label{sec:switches}

In this section, we describe how we implement \name{} to run at line rate on
both network and hypervisor switches. Our implementation assumes that the data
center is running programmable switches like
PISCES~\cite{Shahbaz:2016:PPP:2934872.2934886} and Barefoot
Tofino~\cite{barefoot-tofino}.\footnote{Existing OpenFlow switches may be
configured to simply refer to their group tables when encountering an
\name{} packet. This, however, returns us to the group-table sizes as a
scalability bottleneck.} Having so entails multiple challenges for
programmable switches to efficiently parse, match, and act on {\em
p}-rules.

\subsection{Implementing on Network Switches}
\label{realizing-on-network-switches}
In network switches, typically, a parser first extracts packet headers and then
forwards them to the match-action pipeline for processing. This model works
well for network protocols (like MAC learning and IP routing) that use a
header field to lookup match-action rules in large flow tables. In \name{}, on
the other hand, we find a matching {\em p}-rule from within the packet header
itself. Using match-action tables to perform this matching is prohibitively
expensive (Appendix~\ref{app:lookup-using-tables}). Instead, we present an
efficient implementation by exploiting the {\em match-and-set} capabilities of
parsers in modern programmable data planes.

\paragraph{Matching p-rules using parsers.}
Instead of using a match-action table to lookup {\em p}-rules, the switch can
scan the packet as it arrives at the parser. The parser linearly traverses the
packet header and stores the bits in a header vector based on the configured
parse graph. Parsers in programmable switches provide support for setting
metadata at each stage of the parse
graph~\cite{Bosshart:2013:FMF:2486001.2486011, p4, p4-spec}. Hence, enabling
basic {\em match-and-set} lookups inside the parsers.

\name{} exploits this property, augmenting the parser to check at each
stage---when parsing {\em p}-rules---to see if the identifier of the given {\em
p}-rule matches that of the switch. The parser parses the list of
{\em p}-rules until it reaches a rule with ``next {\em p}-rule'' flag set to 0
(Figure~\ref{fig:header-format}b), or the default {\em p}-rule. If a matching
{\em p}-rule is found, the parser stores the {\em p}-rule's bitmap in a
metadata field and skips parsing remaining {\em p}-rules, jumping directly to
the next header (if any). 

By matching {\em p}-rules inside the parser, we no longer require a
match-action stage to search {\em p}-rules at each switch, thus, making
switch's memory resources available for other use, including {\em s}-rules.
However, the size of a header vector (\ie, the maximum header size a parser can
parse) in programmable switch chips is also fixed. For
RMT~\cite{Bosshart:2013:FMF:2486001.2486011} the size is 512 bytes. We show in
\S\ref{ssec:scalability}, how \name{}'s encoding scheme easily fits enough {\em
p}-rules within 325~bytes while supporting millions of groups. The effective
traffic overhead of 325~bytes is
low~(\S\ref{sssec:flow-table-usage-and-traffic-overhead}), as these {\em
p}-rules get popped with every hop.

\paragraph{Forwarding based on p- and s-rules.}
After parsing the packet, the parser forwards metadata to the ingress pipeline,
which includes a bitmap, a matched flag (indicating the presence of a valid
bitmap), and a default bitmap. The ingress pipeline implements the control
flow to check for the following cases: If the matched flag is set, write the
bitmap metadata to the queue manager~\cite{Bosshart:2013:FMF:2486001.2486011}, 
using a \verb|bitmap_port_select|
primitive~(\S\ref{sssec:enabling-bitmap-based-output-port-selection}). Else,
lookup the group table using the destination IP address for an {\em s}-rule. If
there is a match, write the {\em s}-rule's group identifier to the queue
manager, which then converts it to a bitmap. Otherwise, use the bitmap from the
default {\em p}-rule.

The queue manager generates the desired copies of the packet and forwards
them to the egress pipeline~\cite{Bosshart:2013:FMF:2486001.2486011}. At the
egress pipeline, we execute the following post-processing checks. For leaf
switches, if a packet is going out towards the host, the egress pipeline
invalidates all {\em p}-rules indicating the de-parser to remove these
rules from the packet before forwarding it to the hosts. This offloads the
burden at the receiving hypervisor switches, saving unnecessary CPU cycles
spent to decapsulate {\em p}-rules. Otherwise, the egress pipeline invalidates
all {\em p}-rules up to the {\em p}-rules(s) of the next-hop switch before
forwarding the packet.

\subsection{Implementing on Hypervisor Switches}
\label{realizing-on-hypervisor-switches}
In hardware switches, representing each {\em p}-rule as a separate header is
required to match {\em p}-rules in the parsing stage. However, using
the same approach for the hypervisor switch (like
PISCES~\cite{Shahbaz:2016:PPP:2934872.2934886}) reduces throughput because
each header copy triggers a separate DMA write call. Instead, to operate at
line rate, we treat all {\em p}-rules as one header and encode it using a
single write call (\S\ref{ssec:endhost-microbenchmarks}). Not doing so,
decreases throughput linearly with increasing number of {\em p}-rules to pack.

\begin{table*}[t!]
\centering
\small
\begin{tabular}{>{\raggedright}p{5.5cm}|p{11.5cm}} 
\toprule

\textbf{Hardware resource requirements}: \name{} is inexpensive to implement in
programmable switching ASICs (\S\ref{sec:hardware-results}) &

For a 256-port, 200~$mm^2$ baseline switching ASIC that can parse a 512-byte
packet header~\cite{Gibb:2013:DPP:2537857.2537860}, \name{} consumes only
63.5\% of header space, at the first-hop leaf switch, even with 30 {\em
p}-rules (Figure~\ref{fig:header-sizes}), and its primitives add only 0.0515\%
in area and 176~mW in power costs (Figures~\ref{fig:hardware-mux-area}
and~\ref{fig:hardware-shift-register-area}).
\\ \midrule

\textbf{Scalability}: \name{} scales to millions of multicast groups with
minimal group-table usage and control-plane update overhead on network switches
(\S\ref{ssec:scalability}) &

In a multi-rooted Clos topology having 27K hosts and 1M multicast groups, with
group sizes based on a production trace:

(i) 95-99\% of groups can be encoded using a 325-byte {\em p}-rule header
(Figure~\ref{fig:results-P12} and~\ref{fig:results-P1}, {\em left}).

(ii) Spine and leaf switches use only a mean (max) of 3.8K (11K) and 1.1K
(2.9K) {\em s}-rules (Figure~\ref{fig:results-P12} and~\ref{fig:results-P1},
{\em center}).

(iii) Traffic overhead is kept within 34\% and 5\% of the ideal for 64-byte and
1,500-byte packets, respectively (Figure~\ref{fig:results-P12}
and~\ref{fig:results-P1}, {\em right}).

(iv) On average, a membership change triggers an update to 50\% of hypervisor
switches in a group, less than 0.006\% of leaf and 0.002\% of spine switches
relevant to that group's multicast tree
(Table~\ref{tbl:switch-update-count-normalized}). With 1,000 changes per
second, the average update load on hypervisor, leaf, and spine switches is 21,
5, and 4 updates per second, respectively.
\\ \midrule

\textbf{Applications run unmodified}, and benefit from reduced CPU and
bandwidth utilization for multicast workloads
(\S\ref{sec:application-comparisons}) &

We run ZeroMQ (a publish-subscribe system) and sFlow (a monitoring application)
on top of \name{}. \name{} enables these systems to scale to hundreds of
receivers while maintaining constant CPU and bandwidth overhead at the
transmitting VM (Figure~\ref{fig:zmq-bandwidth-krps-and-cpu-utils}).
\\ \midrule

\textbf{End-host resource requirements}: \name{} adds negligible overheads to
hypervisor switches (\S\ref{ssec:endhost-microbenchmarks}) &  
 
A PISCES-based hypervisor switch encapsulates {\em p}-rules and forwards
packets at line rate on a 20~Gbps link 
(Figure~\ref{fig:hypervisor-adding-headers-throughput-all}). \\ 

\bottomrule
\end{tabular}
\caption{Summary of results.}
\label{tbl:summary-of-results}
\end{table*}

\section{Evaluation}
\label{sec:evaluation}

In this section, we evaluate the resource and scalability requirements of
\name{}. Table~\ref{tbl:summary-of-results} summarizes our results.

\subsection{Hardware Resource Requirements}
\label{sec:hardware-results}
We study the hardware resource requirements of programmable switching ASICs to
process {\em p}-rules. We found \name{} inexpensive to implement in such ASICs.

\subsubsection{Header usage with varying number of {\em p}-rules}
\label{sssec:header-usage}

\begin{figure}[t!]
\centering  
\includegraphics[width=0.85\linewidth]{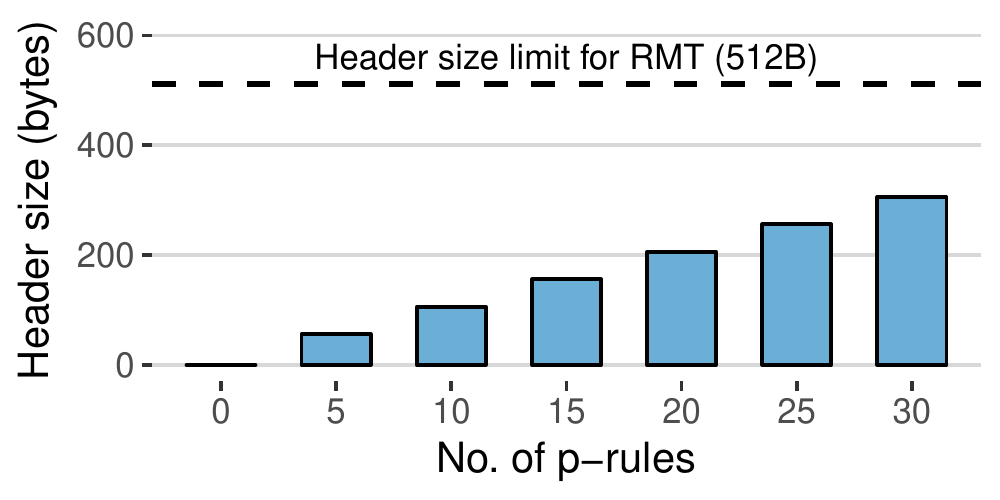}  
\caption{Header usage with varying number of {\em p}-rules, each having four
bytes for switch identifiers and 48 bits for bitmap along with a default {\em
p}-rule. (Horizontal dashed line shows the maximum header space of 512 bytes
for RMT~\cite{Bosshart:2013:FMF:2486001.2486011}.)}
\label{fig:header-sizes}
\end{figure}

Figure~\ref{fig:header-sizes} shows percentage header usage---for a chip that
can parse a 512-byte packet header \eg,
RMT~\cite{Bosshart:2013:FMF:2486001.2486011}---as we increase the number of
{\em p}-rules. Each {\em p}-rule consists of four bytes for switch identifiers
and a 48-bit bitmap. With 30 {\em p}-rules, we are still well within the range,
consuming only 63.5\% of header space for {\em p}-rules with 190 bytes
available for other protocols, which in enterprises~\cite{headers-enterprise}
and
data centers~\cite{headers-datacenter} consume about 90
bytes~\cite{Gibb:2013:DPP:2537857.2537860}. We evaluated these results using
the open-source compiler for P4's behavioral model (\ie, bmv1~\cite{p4c-bm}).

\subsubsection{Enabling bitmap-based output port selection}
\label{sssec:enabling-bitmap-based-output-port-selection}

Data-plane languages (like P4~\cite{p4}) do not yet expose the output-port bit
vector, network switches use for replicating
packets~\cite{Bosshart:2013:FMF:2486001.2486011}, as a metadata field that a
parser can set. We add support for specifying this bit vector using a new
primitive action in P4. We call this new primitive \verb|bitmap_port_select|.
It takes a bitmap of size $N$ as input and sets the output-port bit vector
field that a queue manager then uses to generate copies of a packet, routed to
each egress port. The function is executed by a match-action stage in the
ingress pipeline before forwarding the packet to the queue manager. We evaluate
the primitive using Synopsys 28/32~nm standard cell
technology~\cite{synopsys-technology}, synthesized with a 1~GHz clock. All
configurations meet the timing requirements at 1~GHz.

A typical ASIC implements multicast using a group table that maps a group
identifier to a bit vector~\cite{shankar2003ip, ferolito1999method,
iniewski2008network}. We implement a multiplexer block representing the
hardware requirements to pass this bit vector directly to the queue manager;
and a shift register showing the hardware required to pass the bit vector from
the parser, through all stages of the ingress pipeline, and up to the queue
manager in an RMT-style switching ASIC~\cite{Bosshart:2013:FMF:2486001.2486011,
barefoot-tofino}. We compute the resource requirements for four switch-port
counts: 32, 64, 128, 256.

\begin{figure}[t!]
\centering  
\includegraphics[width=\linewidth]{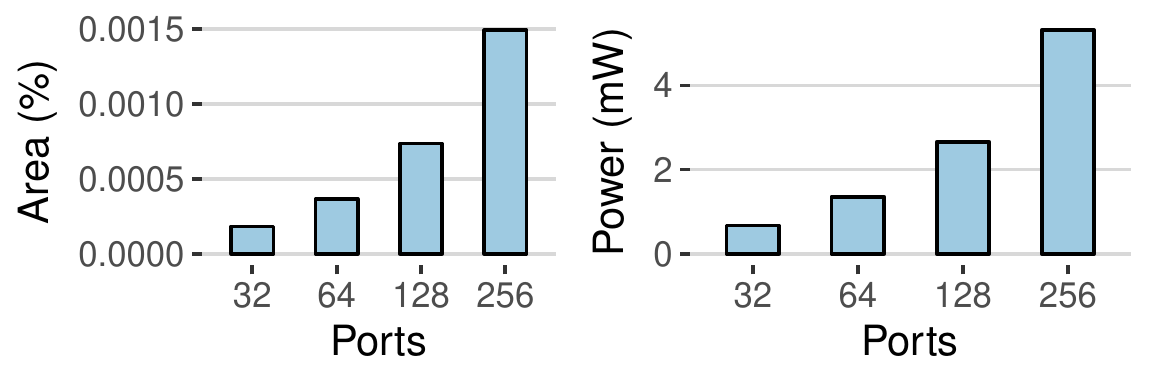}  
\caption{Multiplexer area (in $um^2$) assuming a 200~$mm^2$ area
chip~\cite{Gibb:2013:DPP:2537857.2537860} and power (mW) for different
switch-port counts.}
\label{fig:hardware-mux-area}
\end{figure}

\begin{figure}[t!]
\centering  
\includegraphics[width=\linewidth]
{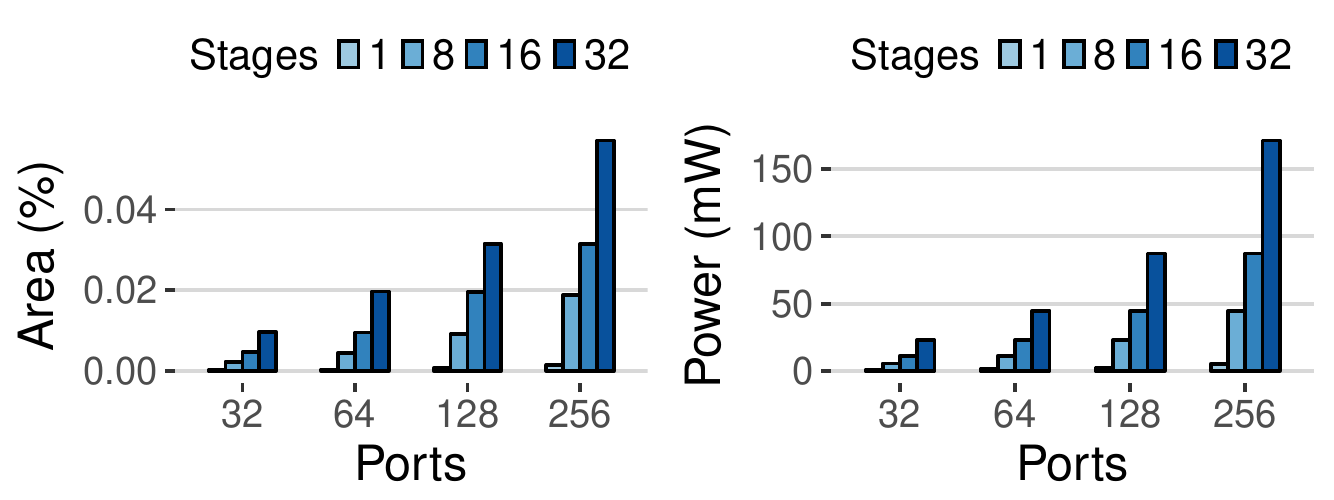}  
\caption{Shift register area (in $um^2$) assuming a 200~$mm^2$ area
chip~\cite{Gibb:2013:DPP:2537857.2537860} and power (mW) for different
switch-port counts.}
\label{fig:hardware-shift-register-area}
\end{figure}

We quantify the area cost and power requirements of adding the multiplexer and
shift register blocks. We evaluate the area cost against a 200~$mm^2$ baseline
switching chip (assuming a lower-end chip and the smallest area given
by~\cite{Gibb:2013:DPP:2537857.2537860}). Figure~\ref{fig:hardware-mux-area}
shows that the added area cost for a multiplexer block is a nominal 0.0015\%
(2,984~$um^2$) for passing a 256-bit wide vector to the queue manager (and a
corresponding 5.31~mW in power). Using a shift register
(Figure~\ref{fig:hardware-shift-register-area}) further increases the area
usage by 0.05\% (114,150.12~$um^2$) for 256-bit wide vectors for a 32 stage
pipeline (and 171.11~mW in power). As another point of comparison for the
reader, CONGA~\cite{Alizadeh:2014:CDC:2619239.2626316} and 
Banzai~\cite{Sivaraman:2016:PTH:2934872.2934900} consume an additional 2\% and
12\% area, respectively.

The circuit delay for the multiplexer is 120~ps (mean) for each of the four
tested port counts. The circuit delay of a single stage in shift register is
150~ps (mean).

\subsection{Scalability}
\label{ssec:scalability}

\subsubsection{Experiment setup}
\label{ssec:experiment-setup}
We now describe the setup we use to test the scale of the number of multicast
groups \name{} can support and the associated traffic and control-plane update
overhead on switches.

\paragraph{Topology.} The scalability evaluation relies on a
simulation over a large data-center topology; the simulation places VMs
belonging to different tenants on end hosts within the data center and assigns
multicast groups of varying sizes to each tenant. We simulate using a Facebook
Fabric topology---a three-tier topology---with 12 pods~\cite{facebook-fabric}.
A pod contains 48 leaf switches each having 48 ports. Thus, the topology with
12 pods supports 27,648 hosts, in total. (We saw qualitatively similar results
while running experiments for a two-tier leaf-spine topology like that used in
CONGA~\cite{Alizadeh:2014:CDC:2619239.2626316}.)

\paragraph{Tenant VMs and placement.} Mimicking the experiment setup from
Multicast DCN~\cite{Li:2013:SIM:2535372.2535380}; the simulated data center has
3,000 tenants; the number of VMs per tenant follows an exponential
distribution, with min=10, median=97, mean=178.77, and max=5,000; and each
host accommodates at most 20 VMs. A tenant's VMs do not share the same physical
host. \name{} is sensitive to the placement of VMs in the data center; which
is typically managed by a placement controller, running alongside the network
controller~\cite{openstack, vmware-vsphere}. We, therefore, perform a
sensitivity analysis using a placement strategy where we first select a pod
uniformly at random, then pick a random leaf within that pod and pack up to
\textbf{\emph{P}} VMs of that tenant under that leaf. \emph{P} regulates the
degree of co-location in the placement. We evaluate for $P=1$ and $P=12$ to
simulate both dispersed and clustered placement strategies. If the chosen leaf
(or pod) does not have any spare capacity to pack additional VMs, the algorithm
selects another leaf (or pod) until all VMs of a tenant are placed.

\paragraph{Multicast groups.} We assign multicast groups to each
tenant such that there are a total of one million groups in the data center.
The number of groups assigned to each tenant is proportional to the size of the
tenant (\ie, number of VMs in that group). We use two different distributions
for groups' sizes, scaled by the tenant's size. Each group's member (\ie, a VM)
is  randomly selected from the VMs of the tenant. The minimum group size is
five. We use the group-size distributions described in the Multicast DCN
paper~\cite{Li:2013:SIM:2535372.2535380}. We model the first distribution 
by analyzing the multicast patterns of an IBM WebSphere Virtual Enterprise
(\emph{\textbf{WVE}}) deployment, with 127 nodes and 1,364 groups. The average
group size is 60, and nearly 80\% of the groups have fewer than 61 members, and
about 0.6\% have more than 700 members. The second distribution generates
tenant's groups' sizes that are uniformly distributed between the minimum group
size and entire tenant size (\emph{\textbf{Uniform}}).

\subsubsection{\name{} scales to millions of multicast groups with
minimal flow-table usage}
\label{sssec:flow-table-usage-and-traffic-overhead}
We first describe results for the various placement strategies under the IBM's
WVE group size distribution. We cap the {\em p}-rule header size at 325 bytes
per packet, which allows up to 30 {\em p}-rules for the downstream leaf layer
and two for the spine layer. We vary the number of redundant transmissions
($R$) permitted due to {\em p}-rule sharing. We evaluate (i)~the number of
groups covered using only the non-default {\em p}-rules, (ii)~the number of
{\em s}-rules installed, and (iii)~the total traffic overhead incurred by
introducing redundancy via {\em p}-rule sharing and default {\em p}-rules.

\begin{figure}[t]
\centering  
\includegraphics[width=\linewidth]{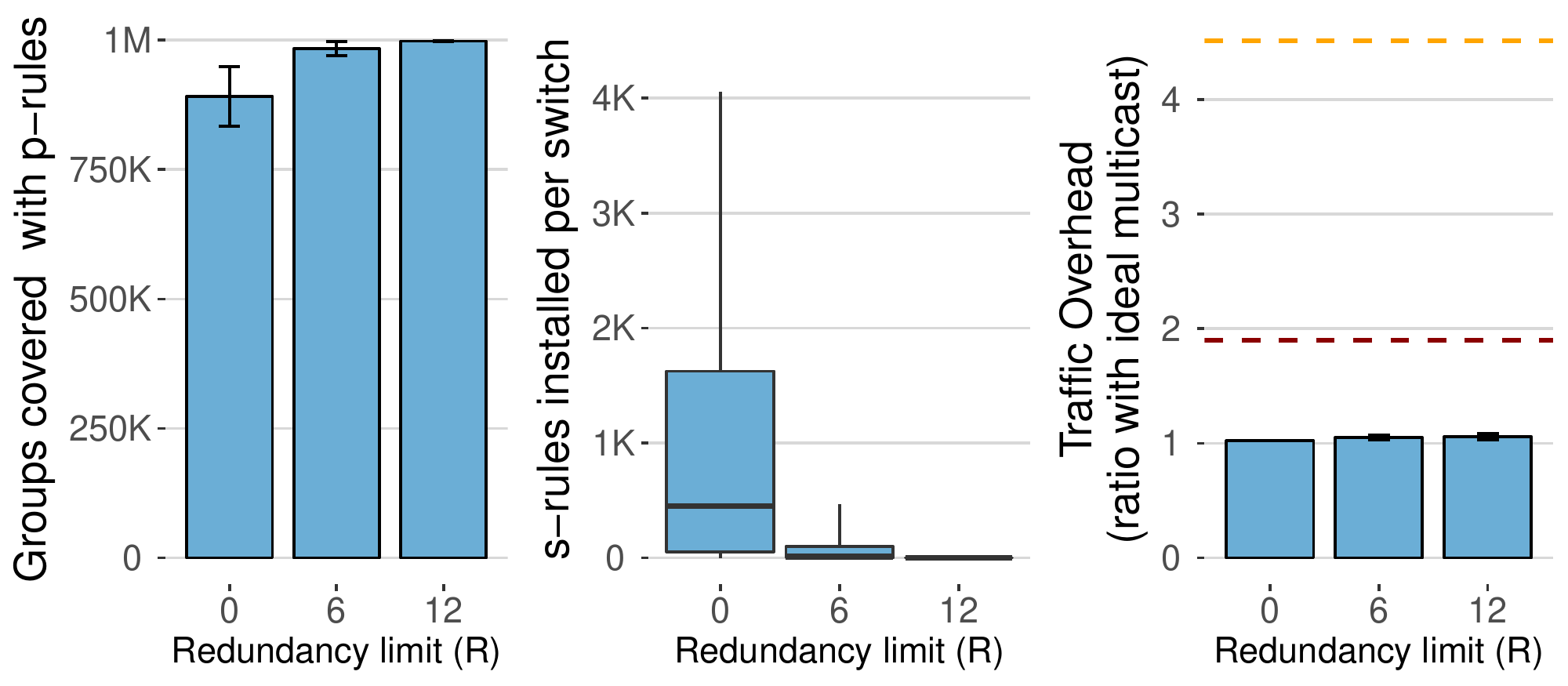}  
\caption{Placement strategy with no more than 12 VMs of a tenant per rack.
\emph{(Left)} Number of groups covered using non-default {\em p}-rules.
\emph{(Center)} {\em s}-rules usage across switches. \emph{(Right)} Traffic
overhead relative to ideal (horizontal dashed lines show unicast (top) and
overlay multicast (bottom)).}
\label{fig:results-P12}
\end{figure}

Figure~\ref{fig:results-P12} shows groups covered with non-default {\em
p}-rules, {\em s}-rules installed per switch, and traffic overhead for a
placement strategy that packs up to 12 VMs of a tenant per rack ($P=12$).
{\em p}-rules suffice to cover a high fraction of groups; 89\% of
groups are covered even when using $R=0$, and 99.78\% with $R=12$. With VMs
packed closer together, the allocated {\em p}-rule header sizes suffice to
encode most multicast trees in the system. Figure~\ref{fig:results-P12}
({\em left}) also shows how increasing the permitted number of extra
transmissions with {\em p}-rule sharing allows more groups to be represented
using only {\em p}-rules.

\begin{figure}[t]
\centering  
\includegraphics[width=\linewidth]{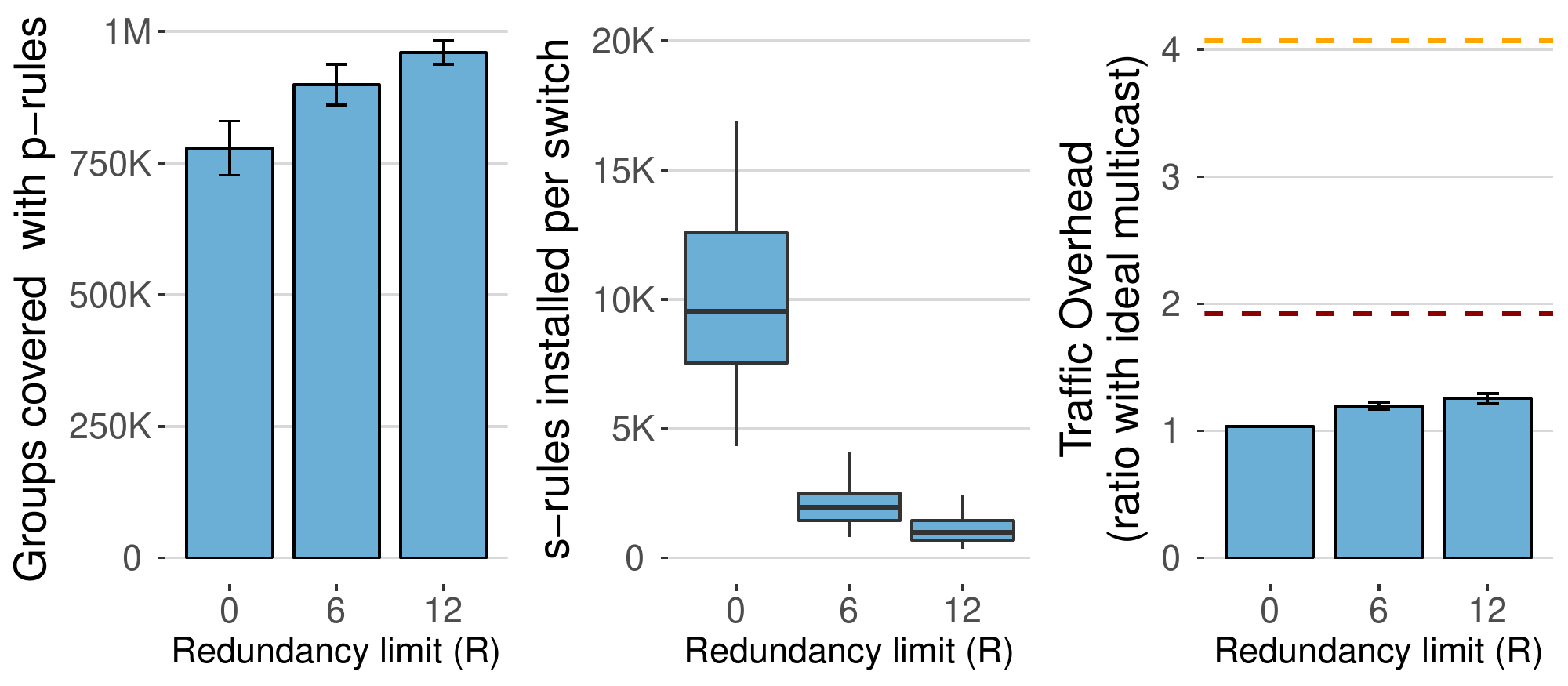}  
\caption{Placement strategy with no more than one VM of a tenant per rack.
\emph{(Left)} Number of groups covered using non-default {\em p}-rules.
\emph{(Center)} {\em s}-rules usage across switches. \emph{(Right)} Traffic
overhead relative to ideal (horizontal dashed lines show unicast (top) and
overlay multicast (bottom)).}
\label{fig:results-P1}
\end{figure}

Figure~\ref{fig:results-P12} ({\em center}) shows the trade-off between {\em
p}-rule and {\em s}-rule usage. With $R=0$, {\em p}-rule sharing tolerates no
redundant traffic. In this case, {\em p}-rules comprise only of switches having
exactly same bitmaps; as a result, the controller must allocate more {\em
s}-rules, with 95\% of leaf switches having fewer than 4,059 rules (mean
1,059). Increasing $R$ to 6 and 12 drastically decreases {\em s}-rule usage as
more groups are handled using only {\em p}-rules. With $R=12$, switches have on
average 2.74 rules, with a maximum of 107.

Figure~\ref{fig:results-P12} ({\em right}) shows the resulting traffic overhead
assuming 1,500-byte packets. With $R=0$ and sufficient {\em s}-rule capacity,
the resulting traffic overhead is identical to ideal multicast. Increasing $R$
increases the overall traffic overhead to 5\% of the ideal. Overhead is modest
because even though a data packet may have as much as 325 bytes of {\em
p}-rules at the source, {\em p}-rules are removed from the header with every
hop (\S\ref{ssec:compact-packet-headers}), reducing the total traffic overhead.
For 64-byte packets, the traffic overhead for WVE increases only to 29\% and
34\% of the ideal when $R=0$ and $R=12$, still significantly improving over
overlay multicast\footnote{In overlay multicast, the source host's hypervisor
switch replicates packets to one host under each participating leaf switch,
which then replicates packets to other hosts under that leaf switch.} (92\%)
and unicast (406\%).

\paragraph{p-rule sharing is effective even when groups are dispersed across
leaves.} Thus far, we discussed results for when up to 12 VMs
of the same tenant were placed in the same rack. To understand how our results
vary for different VM placement strategies, we explore an extreme case where
the placement strategy spreads VMs across leaves, placing no more than a single
VM of a tenant per rack. Figure~\ref{fig:results-P1} ({\em left}) shows this
effect. Dispersing groups across leaves requires larger headers to encode the
whole multicast tree using only {\em p}-rules. Even in this case, {\em p}-rules
with $R=0$ can handle as many as 750K groups, since 77.8\% of groups have less
than 36 switches, and there are 30 {\em p}-rules for the leaf layer---just
enough header capacity to be covered only with {\em p}-rules.  
Increasing $R$ to 12 ensures that 95.9\% of groups are covered using
{\em p}-rules. We see the expected drop in {\em s}-rule usage as well, in
Figure~\ref{fig:results-P1} ({\em center}), with 95\% of switches having fewer
than 2,435 {\em s}-rules. The traffic overhead increases to within 25\% of the
ideal when $R=12$, in Figure~\ref{fig:results-P1} ({\em right}), but still
improving significantly over overlay multicast (92\%) and unicast (406\%).

\paragraph{p-rule sharing is robust to different group size distributions.} We
also study how the results are affected by different distributions of group
sizes, using the Uniform group size distribution. We expect that larger group
sizes will be more difficult to encode using only {\em p}-rules. We found that
with the $P=12$ placement strategy, the total number of groups covered using
only {\em p}-rules drops to 814K at $R=0$ and to 922K at $R=12$. When spreading
VMs across racks with $P=1$, only 250K groups are covered by {\em p}-rules
using $R=0$, and 750K when $R=12$. The total traffic overhead for 1,500-byte
packets in that scenario increases to 11\%.

\paragraph{Reducing s-rule capacity increases default p-rule usage if p-rule
sizes are insufficient.} Limiting the {\em s}-rule capacity of switches
allows us to study the effects of limited switch memory on the efficiency of
the encoding scheme. Doing so increases the number of switches that are mapped
to the default {\em p}-rule. When limiting the {\em s}-rules per switch to 10K
rules, and using the extreme $P=1$ placement strategy, the uniform group size
distribution experiences higher traffic overheads, approaching that of overlay
multicast at $R=0$ (87\% vs 92\%), but still being only 40\% over ideal
multicast at $R=12$. Using the WVE distribution, however, brings down traffic
overhead to 19\% and 25\% for $R=6$ and $R=12$, respectively. With the tighter
placement of $P=12$,  however, we found the traffic overhead to consistently
stay under 5\% regardless of the group-size distribution.

\paragraph{Reduced p-rule header sizes and s-rule capacities inflate traffic
overheads.} Finally, to study the effects of the size of the {\em p}-rule
header, we reduced the size so that the header could support at most 10 {\em
p}-rules for the leaf layer (\ie, 125 bytes per header). In conjunction, we
also reduced the {\em s}-rule capacity of each switch to 10K and used the $P=1$
placement strategy to test a scenario with maximum dispersement of VMs. This
challenging scenario even brought the traffic overhead to exceed that of
overlay multicast at $R=12$ (123\%). However, in contrast to overlay
multicast, \name{} still forwards packets at line rate without any overhead on
the end host CPU utilization.

\subsubsection{\name{} is robust to membership churn and network failures}
\label{sssec:update-overhead}

We use the same Facebook Fabric setup to evaluate the effects of
group membership churn and network failures on the control-plane update
overhead on switches.

\paragraph{Group membership dynamics.} 
In \name{}, we distinguish between three types of members: senders, receivers,
or both. For this evaluation, we randomly assign one of these three types to
each member. All VMs of a tenant who are not a member of a group have equal
probability to join; similarly, all existing members of the group have an equal
probability of leaving. Join and leave events are generated randomly, and the
number of events per group is proportional to the group size. 

If a member is a sender, the controller only updates the source hypervisor
switch. By design, \name{} only uses {\em s}-rules if the {\em p}-rule header
capacity is insufficient to encode the entire multicast tree of a group.
Membership changes trigger updates to sender and receiver hypervisor switches
of the group depending on whether upstream or downstream {\em p}-rules need to
be updated. When a change affects {\em s}-rules, it triggers updates to the
leaf and spine switches.

\begin{table}[t!]
\centering
\small
\begin{tabular}{ p{1.8cm} p{1.5cm} | c }
\toprule
\textbf{switch} & \textbf{event} &
\textbf{$\frac{\boldsymbol{\#~Switches~updated~per~event}}{\boldsymbol{Group
~Size}}$}
\\
\midrule
\multirow{2}{*}{hypervisor} & join & 0.3351 \\
                            & leave & 0.4999 \\
                            
\hline
\multirow{2}{*}{leaf} & join & 0.0042 \\ 
                      & leave & 0.0061 \\
\hline
\multirow{2}{*}{spine} & join & 0.0015 \\
                       & leave & 0.0023 \\ 
\bottomrule
\end{tabular}
\caption{The average number of hypervisor, leaf, and spine switches updated
during an event in a multicast tree of a group (normalized by group sizes).
Results are shown for WVE distribution.}
\label{tbl:switch-update-count-normalized}
\end{table}

Table~\ref{tbl:switch-update-count-normalized} shows the results for one
million join/leave events with one million multicast groups, where no more than
one VM of a tenant is placed per rack. On average, a membership change 
triggers an update to 50\% (max 2,588) of hypervisor switches in a group, fewer
than 0.006\% (max 35) of leaf and 0.002\% (max 24) of spine switches relevant
to that group's multicast tree, demonstrating that hypervisor switches handle
most of \name{}'s control-plane updates. Cloud management platforms (like
OpenStack~\cite{openstack}), today,
can easily handle concurrent updates to 10K hypervisor switches and up to 1,000
network switches~\cite{vmware-vsphere-configs,
NiranjanMysore:2009:PSF:1592568.1592575}. 

The update load (per second) on these switches also remains well within the
studied thresholds~\cite{Huang:2013:HSM:2491185.2491188}; with membership
changes of 1,000 events per second, the average (max) update load on
hypervisor, leaf, and spine switches is 21 (46), 5 (13), and 4 (7) updates per
second, respectively. Hypervisor and network switches can support up to 2,000
and 100 updates per second~\cite{ovs-perf, He:2015:MCP:2774993.2775069},
implying that we can support a demand of 44K and 8K membership changes per
second, respectively, before hitting the limit of these switches.

\paragraph{Network failures.} \name{} gracefully handles spine and core
switch failures,\footnote{When a leaf switch fails, all hosts connected to it
lose connectivity to the network, and must wait for the switch to get
back online.} forwarding multicast packets via alternate paths using upstream
ports represented in the groups' {\em p}-rule
bitmap~(\S\ref{sub:upstream-ports}). During this period, some groups might
experience transient loss while the network is
reconfiguring~\cite{NiranjanMysore:2009:PSF:1592568.1592575}. In our
simulations, up to 12.25\% of groups are impacted when a single spine
switch fails and up to 25.81\% when a core switch fails. Hypervisor
switches incur average (max) updates of 176.86 (1712) and 674.89 (1852),
respectively. We measure that today's hypervisor switches are capable of
handling {\em batched} updates of 80K requests per second (on a modest server)
and, hence, can reconfigure within 25~ms of these failures.

\subsubsection{\name{}'s controller computes {\em p}- and {\em s}-rules for a
group within a millisecond}
\label{sec:processing-time} 

Our controller consistently executes Algorithm~\ref{alg:cluster} for computing
{\em p}- and {\em s}-rules in less than a millisecond. Across our simulations,
our Python implementation computes the required rules for each group in
0.20~ms $\pm$ 0.45~ms (SD), on average, for all group sizes with a header
size limit of 325 bytes. Existing studies report that it takes up to 100~ms
for a controller to learn an event, issue updates to the network, and for the
network state to converge~\cite{NiranjanMysore:2009:PSF:1592568.1592575}.
\name{}'s control logic, therefore, contributes little to the overall
convergence time for updates and is fast enough to support the needs of large
data centers today, even before extensive optimization. 

\subsection{Evaluating End-to-end Applications}
\label{sec:application-comparisons}

We ran two popular data-center applications on top of \name{}:
ZeroMQ~\cite{hintjens2013zeromq} and sFlow~\cite{rfc3176}. We found that these
applications ran unmodified on top of \name{} and benefited from reduced CPU
and bandwidth utilization for multicast workloads.

\paragraph{Testbed setup.}
The topology for this experiment comprises nine PowerEdge R620 servers
having two eight cores Intel(R) Xeon(R) CPUs running at 2.00~GHz and with 32~GB
of memory, and three dual-port Intel 82599ES 10 Gigabit Ethernet NICs. Three
of these machines emulate a spine and two leaf switches; these machines run an 
extended version of the PISCES~\cite{Shahbaz:2016:PPP:2934872.2934886} switch
with
support for the \verb|bitmap_port_select| primitive for routing 
traffic between interfaces. The remaining machines act as hosts, with three 
hosts per leaf switch.

\begin{figure}[t!]
\centering  
\includegraphics[width=\linewidth]
{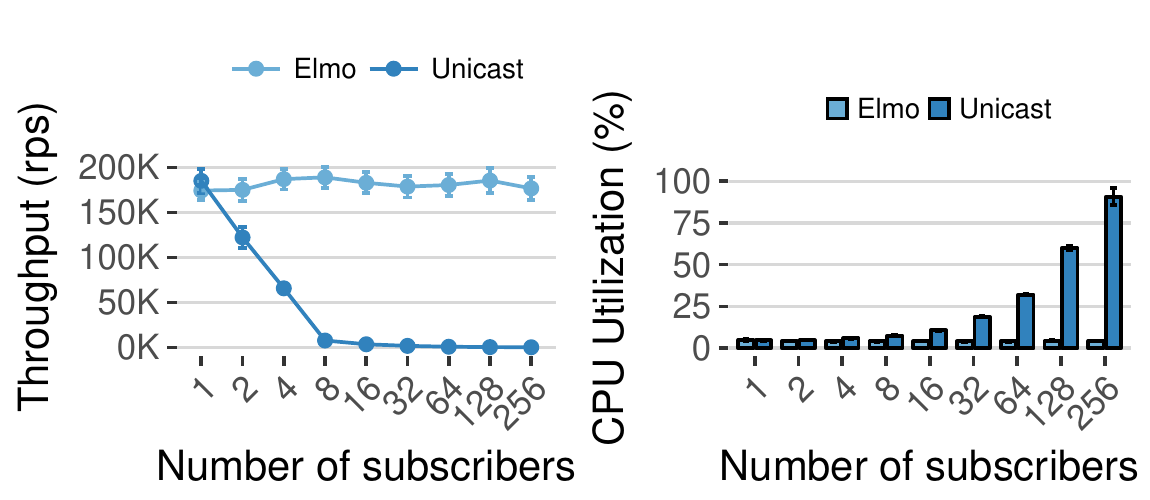}
\caption{Requests-per-second and CPU utilization comparison of a pub-sub
application using ZeroMQ (over UDP) for both unicast and \name{} with a 
message size of 100~bytes.}
\label{fig:zmq-bandwidth-krps-and-cpu-utils}
\end{figure}

\subsubsection{Publish-subscribe using ZeroMQ}
\label{sssec:pubsub-using-zeromq}
We implement a publish-subscribe (pub-sub) system using ZeroMQ (over UDP).
ZeroMQ enables tenants to build pub-sub systems on top of a cloud environment
(like AWS~\cite{amazon-aws}, GCP~\cite{google-gcp}, or
Azure~\cite{microsoft-azure}), by establishing unicast connections between
publishers and subscribers.\footnote{A drawback of native multicast is that we
cannot use TCP. However, protocols (like PGM~\cite{rfc3208} and
SRM~\cite{Floyd:1995:RMF:217382.217470}) can support applications that require
reliable delivery using native multicast.}

\paragraph{Throughput (rps).}
Figure~\ref{fig:zmq-bandwidth-krps-and-cpu-utils} ({\em left}) shows the
throughput comparison in requests per second. With unicast, the throughput at
subscribers decreases with an increasing number of subscribers because the
publisher becomes the bottleneck; the publisher services a single subscriber at
185K~rps on average and drops to about 0.25K~rps for 256 subscribers. With
\name{}, the throughput remains the same regardless of the number of
subscribers and averages 185K~rps throughout.

\paragraph{CPU utilization.} The CPU usage of the publisher VM (and the
underlying host) also increases with increasing number of subscribers,
Figure~\ref{fig:zmq-bandwidth-krps-and-cpu-utils} ({\em right}). The publisher
VM consumes 32\% of the VM's CPU with 64 subscribers and saturates the CPU with
256 subscribers onwards. With \name{}, the CPU usage remains constant
regardless of the number of subscribers (\ie, 4.97\%).

\subsubsection{Host telemetry using sFlow}
\label{sssec:host-telemetry-using-sflow}
As our second application, we compare the performance of host telemetry using
sFlow with both unicast and \name{}. sFlow exports physical and virtual server
performance metrics from sFlow agents to collector nodes (\eg, CPU, memory, and
network stats for docker, KVMs, and hosts) set up by different tenants (and
teams) to collect metrics for their needs. We compare the egress bandwidth
utilization at the host of the sFlow agent with increasing number of
collectors, using both unicast and \name{}. The bandwidth utilization increases
linearly with unicast, with the addition of each new collector. With 64
collectors, the egress bandwidth utilization at the agent's host is
370.35~Kbps. With \name{}, the utilization remains constant at about 5.8~Kbps
(equal to the bandwidth requirements for a single collector).

\subsection{End-host Microbenchmarks}
\label{ssec:endhost-microbenchmarks}
We conduct microbenchmarks to measure the incurred overheads on the hypervisor
switches when encapsulating {\em p}-rule headers onto packets (decapsulation at
every layer is performed by network switches). We found \name{} imposes
negligible overheads at hypervisor switches.

\paragraph{Setup.} Our testbed has a host $H_1$ directly connected to two hosts
$H_2$ and $H_3$. $H_1$ has 20~Gbps connectivity with both $H_2$ and $H_3$, via
two 10~Gbps interfaces per host. $H_2$ is a traffic source and $H_3$ is a
traffic sink; $H_1$ is running PISCES with the extensions for \name{} to
perform necessary forwarding. $H_2$ and $H_3$ use
MoonGen~\cite{Emmerich:2015:MSH:2815675.2815692} for generating and receiving
traffic, respectively.

\begin{figure}[t!]
\centering  
\includegraphics[width=\linewidth]
{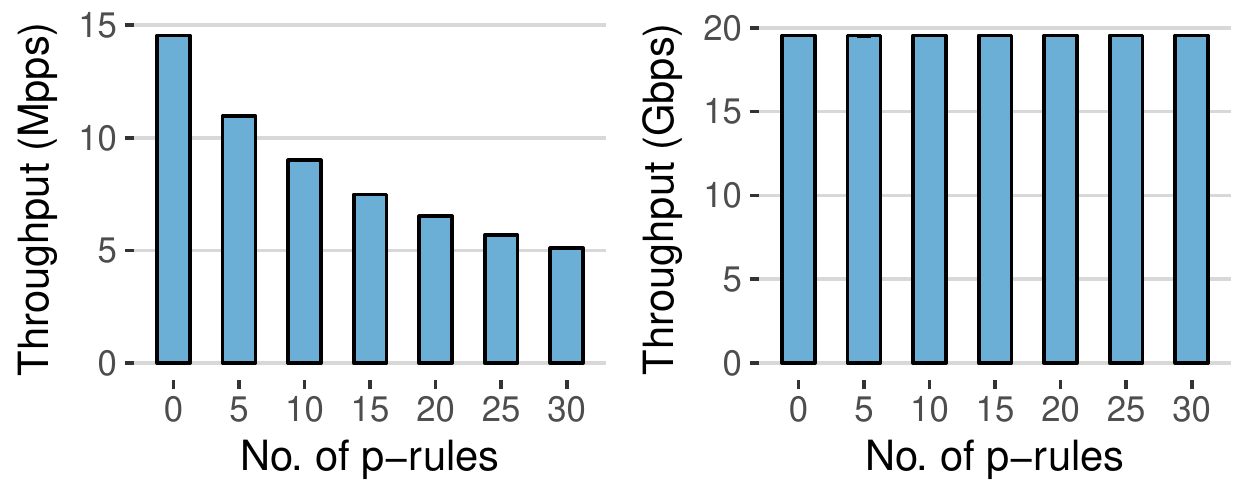}  
\caption{PISCES throughput in millions of packets per second (\emph{left})
and Gbps (\emph{right}) when adding different number of {\em p}-rules,
expressed as a single P4 header.}
\label{fig:hypervisor-adding-headers-throughput-all}
\end{figure}

\paragraph{Results.}
Figure~\ref{fig:hypervisor-adding-headers-throughput-all}
shows throughput at a hypervisor switch when encapsulating different number
of {\em p}-rule headers, in both packets per second (pps) and Gigabits per
second (Gbps). Increasing the number of {\em p}-rules reduces the pps rate, as
the packet size increases, while the throughput in bps remains unchanged. The
throughput matches the capacity of the links at 20~Gbps, demonstrating that
\name{} imposes negligible overhead on hypervisor switches.

\section{Related Work}\label{sec:related-work}

\paragraph{Wide-area multicast.} Multicast has been studied in detail in the
context of wide-area networks
~\cite{Deering:1994:AWM:190314.190326, Costa:2001:HHM:383059.383079,
Crowcroft:1988:MTP:52324.52349, Samadi:2014:AIM:2619239.2631436,
Ballardie:1993:CBT:166237.166246}, where the lack of applications and
architectural complexities led to limited
adoption~\cite{Diot:2000:DII:2329003.2329353}. Furthermore, the decentralized
protocols such as IGMP and PIM faced several control-plane challenges with
regards to stability in the face of membership
churn~\cite{Diot:2000:DII:2329003.2329353}. Over the years, much work has gone
into IP multicast to address issues related to
scalability~\cite{cisco-scalability, Jokela:2009:LLS:1592568.1592592},
reliability~\cite{Balakrishnan:2007:RLE:1973430.1973436,
balakrishnan2005slingshot, Floyd:1995:RMF:217382.217470, lehman1998active},
security~\cite{judge2003security}, and congestion
control~\cite{Widmer:2001:EEC:383059.383081, rfc4654}. \name{}, however,
is designed for data centers which differ in significant ways from the
wide-area context.

\paragraph{Data-center multicast.}  With SDN-based data centers, a
single administrative domain has control over the entire topology and is no
longer required to run the decentralized protocols like IGMP and PIM. However,
SDN-based multicast is still bottlenecked by limited switch group-table
capacities~\cite{hardware-group-limit, multicast-group-capacity,
multicast-sdn-vxlan}. Approaches to scaling multicast groups in this context
have tried using rule aggregation to share multicast entries in switches with
multiple groups~\cite{Li:2013:SIM:2535372.2535380,
Lin:2017:SMM:3031106.3031152, cui2014dual, iyer2014avalanche}. Yet, these
solutions do not operate well in cloud environments because (1) a change in one
group can cascade to other groups, (2) they do not provide address-space
isolation between tenants, and (3) they cannot utilize the full bisection
bandwidth of the network~\cite{Li:2013:SIM:2535372.2535380,
NiranjanMysore:2009:PSF:1592568.1592575}. \name{}, on the other hand, operates
on a group-by-group basis, maintains address-space isolation, and makes full
use of the entire bisection bandwidth.

\paragraph{Application/overlay multicast.} The lack of IP multicast support,
including among the major cloud providers~\cite{aws-faq, gcp-faq,
azure-faq}, requires tenants to use inefficient software-based multicast
solutions such as overlay multicast or application-layer
mechanisms~\cite{hintjens2013zeromq, videla2012rabbitmq,
Banerjee:2002:SAL:633025.633045, Das:2002:SSW:647883.738420, consul-faq}. These
mechanisms are built on top of unicast, which as we demonstrated in
\S\ref{sec:evaluation}, incurs a significant reduction in application
throughput and inflates CPU utilization. SmartNICs (like Netronome's
Agilio~\cite{netronome-agilio}) can help offload packet-replication burden from
end hosts' CPUs. However, these NICs are limited in their capabilities (such as
flow-table sizes and the number of packets they can clone). The replicated
packets contend for the same egress port; further restricting these NICs from
sustaining line rate and predictable latencies. With native multicast, as in
\name{}, end hosts send a single copy of the packet to the network and use
intermediate switches to replicate and forward copies to multiple destinations
at line rate.

\paragraph{Source-routed multicast.} \name{} is not the first system to encode
forwarding state inside packets. Previous
work~\cite{Ratnasamy:2006:RIM:1159913.1159917, Jokela:2009:LLS:1592568.1592592,
li2011scalable} have tried to encode link identifiers inside packets using
bloom filters. BIER~\cite{rfc8279} encodes group members as bit strings,
whereas SGM~\cite{boivie2000small} encodes them as a list of IP addresses.
Switches then look up these encodings to identify output ports. However, all
these approaches require unorthodox processing at switches (\eg, loops,
multiple lookups on a single table, and more) and are infeasible to implement
and process multicast traffic at line rate. BIER, for example, requires flow
tables to return all entries (wildcard) matching the bit strings---a
prohibitively expensive data structure compared to traditional match-action
tables in emerging programmable data planes. SGM, on the other hand, looks up
all the IP addresses in the routing table to find their respective next hops,
requiring an arbitrary number of routing table lookups, thus, breaking the
line-rate invariant. Contrary to these approaches, \name{} is designed to
operate at line rate using modern programmable data planes (like Barefoot
Tofino~\cite{barefoot-tofino} and Cavium XPliant~\cite{xpliant}).

\balance\section{Conclusion}
\label{sec:conclusion}

In this paper, we presented \name{}, a solution to scale multicast to millions
of groups per data center. \name{} encodes multicast forwarding rules inside
packets themselves, reducing the need to install corresponding group-table
entries in network switches. \name{} takes advantage of emerging programmable
switches and unique characteristics of data-center topologies to identify
compact encodings of multicast forwarding rules inside packets. Our simulations
show that a 325-byte header sufficed to support a million multicast groups in a
three-tier data center with 27K hosts, while using minimal group-table entries
in network switches. Furthermore, \name{} is inexpensive to implement in
programmable switches today and supports applications that use multicast
without modification.
\label{lastpage}

{\footnotesize \bibliographystyle{acm}
\bibliography{paper}}


\appendix
\section{Strawman: {\em p}-rule lookups using match-action stages is expensive}
\label{app:lookup-using-tables}

Lookups in network switches are typically done using match-action tables, after
the parser. We could do the same for {\em p}-rules, but using match-action
tables to lookup {\em p}-rules would result in inefficient use of switch
resources. Unlike {\em s}-rules, {\em p}-rules are headers. Hence, to match on
{\em p}-rules, we need a table that matches on all {\em p}-rule headers. In
each flow rule, we only match the switch identifier with one {\em p}-rule,
while wildcarding the rest. This is a constraint of match-action tables in
switches that we cannot avoid. To match $N$ {\em p}-rules, we need same number
of flow-table entries.

The fundamental problem here is that instead of increasing the \emph{depth},
{\em p}-rules increase the \emph{width} of a table. Modern programmable 
switches can store millions of flow-table entries (depth). However, they are
severely limited by the number of headers they can match on in a stage (width).
For example, in case of RMT~\cite{Bosshart:2013:FMF:2486001.2486011}, a
match-action stage consists of 106 1K x 112b SRAM blocks and 16 2K x 40b TCAM
blocks. These blocks can combine together to build wider or deeper SRAMs
and TCAMs to make larger tables. For example, to implement a table that matches
on ten {\em p}-rules, each 11-bit wide, we need three TCAM blocks (as we need
wildcards) to cover 110b for the match. This results in a table of 2K entries x
120b wide. And out of these 2K entries, we only use ten entries to match the
respective {\em p}-rules. Thus, we end up using three TCAMs for ten {\em
p}-rules while consuming only 0.5\% of entries in the table, wasting 99.5\% of
the entries (which cannot be used by any other stage).

An alternative to using TCAMs for {\em p}-rule lookups is to eschew wildcard
lookups and use SRAM blocks. In this case, a switch needs $N$ stages to lookup
$N$ {\em p}-rules in a packet, where each stage only has a single rule. This
too is prohibitively expensive. First, the number of stages in a switch is
limited (RMT has 16 stages for the ingress pipeline). Second, as with TCAMs,
99.9\% of the SRAM entries go to waste, as each SRAM block is now used only for
a single {\em p}-rule each (out of 1K available entries per block).

\end{document}